\documentclass[final, 5p]{elsarticle}

\usepackage[section]{placeins}
\usepackage{amsmath}
\usepackage{amsfonts}
\usepackage{multirow}
\usepackage{verbatim}
\usepackage{color}     
\usepackage{array}
\usepackage{dcolumn}
\usepackage{xspace}
\usepackage{textcomp}
\usepackage{graphicx}
\usepackage{gensymb}
\usepackage{textgreek}
\usepackage[colorlinks]{hyperref}
\usepackage{amssymb}
\usepackage[export]{adjustbox}
\usepackage{esvect}

\journal{Nuclear Instruments and Methods in Physics Research A}

\begin{document}

\title{Metastability exchange optical pumping of $^3$He at low pressure and high magnetic field}

\author[MIT]{X.~Li}\corref{cor1}\ead{xqli@mit.edu}
\author[JLab]{J.~D.~Maxwell}
\author[JLab]{D.~Nguyen}
\author[JLab]{J.~Brock}
\author[JLab]{C.~D.~Keith}
\author[MIT]{R.~G.~Milner}
\author[JLab]{X.~Wei}

\affiliation[MIT]{organization={Laboratory for Nuclear Science, Massachusetts Institute of Technology},
            city={Cambridge},
            state={MA 02139},
            country={USA}}
\affiliation[JLab]{organization={Thomas Jefferson National Accelerator Facility},
            city={Newport News},
            state={VA 23606},
            country={USA}}
\cortext[cor1]{Corresponding author.}

\begin{abstract}
Systematic studies on metastability exchange optical pumping of $^3$He nuclei have been performed at Jefferson Lab using a 1-torr sealed cell at magnetic fields from 2 to 4\,T. The effects of the discharge intensity, pump laser power and pumping transition schemes on achievable nuclear polarization and pumping rate have been investigated. A maximum steady-state nuclear polarization of about 75\% has been obtained. This work provides a baseline for the development of the novel polarized $^3$He target for CLAS12 at Jefferson Lab.
\end{abstract}

\begin{keyword}

Polarized helium-3
\sep
Metastability exchange optical pumping
\sep
High magnetic field

\end{keyword}

\maketitle
\date{}

\section{Introduction}
\label{sec:intro}
Nuclear spin-polarized $^3$He is a powerful effective polarized neutron target which plays a significant role in the studies on neutron spin structure. During the last three decades, spin-polarized $^3$He gas targets have been well developed using optical pumping techniques~\cite{Colegrove:1960,Bouchiat:1960dsd} and successfully implemented in scattering experiments at MIT-Bates~\cite{Jones:1993hg}, SLAC~\cite{Johnson:1994cq}, DESY~\cite{DeSchepper:1998gc}, Mainz~\cite{Krimmer:2009zz}, HIGS~\cite{Kramer:2007zzb} and JLab~\cite{Singh:2010} to study the fundamental quark and gluon dynamics inside the nucleon and nucleus. The metastability exchange optical pumping (MEOP) technique utilizes 1083-nm circularly polarized laser light to produce nuclear polarization in metastable-state $^3$He atoms via optical pumping and hyperfine coupling. The polarization is then transferred to the ground-state $^3$He nuclei through metastability-exchange collisions. The double-cell polarized $^3$He target designed and built by Caltech~\cite{Milner:1989} was the first MEOP target developed for electron scattering experiments, and was used in the MIT-Bates 88-02 experiment to carry out the first measurement of spin-dependent electron scattering from the $^3$He nucleus. In this target, $^3$He was polarized at low field using the MEOP technique in a room-temperature pumping cell and then diffused into a cold (below 20\,K) target cell to enhance the gas density as MEOP takes place in low pressures of a few mbar. Typical in-beam nuclear polarizations of 30--40\% and a target thickness of $\sim$$10^{19}$ nuclei/cm$^2$ had been proven without significant depolarization from the relativistic electron beam currents up to 40 μA~\cite{Jones:1993hg,Gao:1994ud}. The effects of gas pressure, temperature and wall relaxation had also been systematically studied~\cite{Jones thesis, Gao thesis}.

Traditional polarized $^3$He targets operate well in low magnetic fields (on the order of 10$^{-3}$\,T) but have not been extensively explored in high-magnetic-field environments. The recent development on high-field MEOP of $^3$He ~\cite{Courtade:2000, Courtade:2002,Abboud:2004,Abboud:2005,Nikiel:2007,Suchanek:2007,Nikiel-Osuchowska:2013} has opened up new opportunities to extend the application of polarized $^3$He targets to experimental facilities involving high magnetic fields, such as the Electron Beam Ion Source (EBIS) at the Brookhaven National Lab (BNL) and the CLAS12 spectrometer at JLab. At BNL, development on a polarized $^3$He ion source within the EBIS 5\,T solenoid is currently underway for the future Electron-Ion Collider~\cite{Zelenski:2023kof}. At JLab, development of a novel polarized $^3$He target for CLAS12 has been motivated by the success of the MIT-Bates target and the high-field MEOP studies at BNL~\cite{Maxwell:2018dyf}. The conceptual design for the CLAS12 polarized $^3$He target~\cite{Maxwell:2021ytu} combines the double-cell cryogenic design of the MIT-Bates target and the recently developed high-field MEOP technique, aiming to produce polarized $^3$He inside the CLAS12 5\,T solenoid with a target thickness comparable to traditional low-field polarized $^3$He targets. In this way, the full effectiveness of the state-of-the-art CLAS12 detector can be utilized in measurements of spin-dependent electron scattering from polarized $^3$He across the complete kinematic range, {\it e.g}, elastic, quasielastic, resonance, deep inelastic, deeply virtual exclusive, etc.  Such a target is motivated by the conditionally approved proposal~\cite{PAC:2020} to measure spin-dependent deep inelastic electron scattering from polarized $^3$He at CLAS12 in Hall B at JLab. Recently, a high-field MEOP system for polarized $^3$He has been established at JLab to initiate this target R\&D work. Systematic studies on the effects of the discharge intensity, pump laser power and optical pumping transition schemes on $^3$He nuclear polarization have been conducted. In this work, we report the major findings from these systematic studies.

\section{High-field MEOP}
\label{sec:meop}
The production of $^3$He nuclear polarization using MEOP involves optical pumping of $^3$He in the metastable state and metastability-exchange collisions. A radio-frequency (RF) signal is employed to induce electrical plasma discharge in $^3$He gas and excite a small population of $^3$He atoms from the ground state to the $2^3S_1$ metastable state. The $2^3S_1$--$2^3P$ optical pumping transition is then driven by circularly polarized 1083-nm laser light. Atoms in the $2^3P$ state are brought back to the $2^3S_1$ state by spontaneous or stimulated emissions. The optical pumping process gives rise to the electronic polarization in the metastable-state $^3$He atoms, which is then partially passed to the $^3$He nuclei by the hyperfine interaction. Finally the nuclear polarization in the metastable-state $^3$He is transferred to the ground-state $^3$He via metastability-exchange collisions.

\begin{figure}[htp]
\begin{center}
\includegraphics[width=1.2\columnwidth,center]{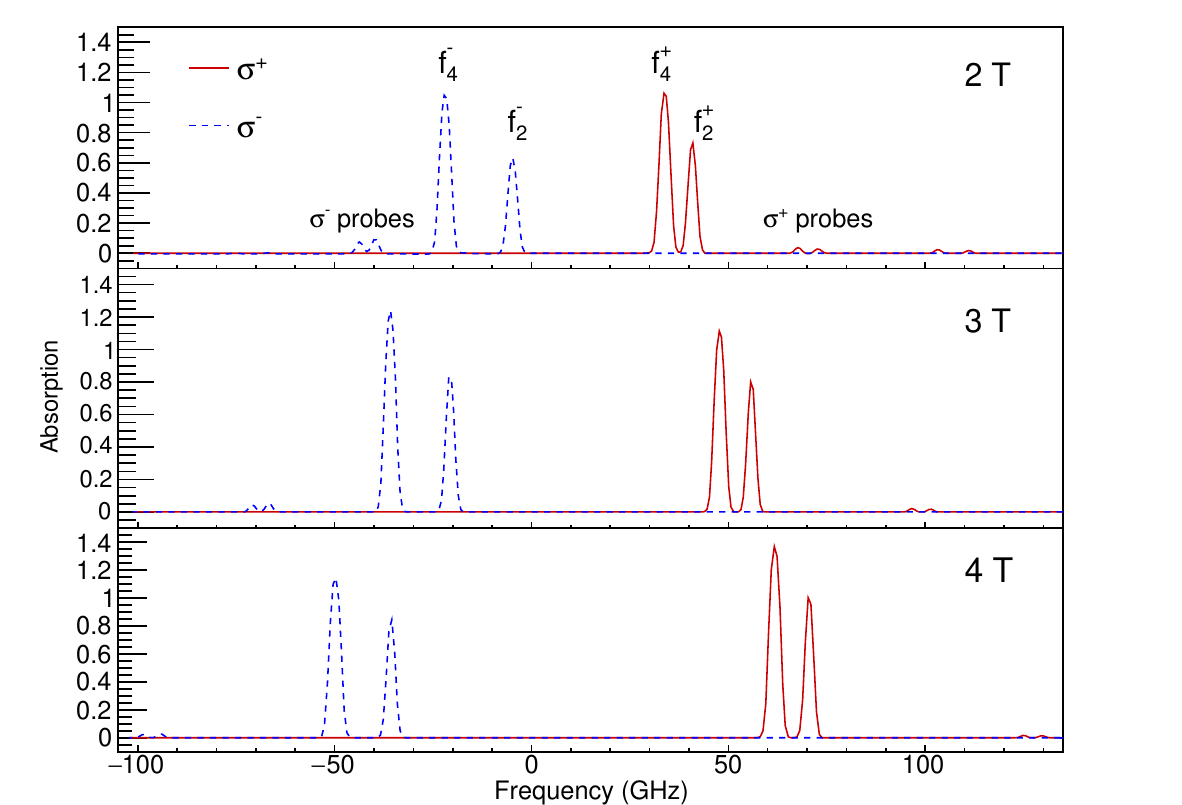}
\caption{Measured absorption spectra for the 1083-nm optical pumping and probing transitions for $^3$He at 2, 3 and 4\,T. The frequency is relative to the $C_1$ transition ($2^3S_1,F=\frac{1}{2} \rightarrow 2^3P,F=\frac{3}{2}$) at B = 0.
\label{fig:peaks}}
\end{center}
\end{figure}

The Zeeman sublevels of the $2^3S_1$ and $2^3P$ states of $^3$He significantly differ between low and high magnetic fields, and thence the 1083-nm optical pumping transitions. This results in different optical pumping and polarimetry approaches for low- and high-field MEOP. In a low field, the $C_8$ and $C_9$ transition lines are adopted to promote the metastable-state $^3$He to the $2^3P$ state (see Fig.~14 in Ref.~\cite{Gentile:2016uud}) and the nuclear polarization of $^3$He can be measured by observing the circular polarization of the 668-nm light emitted by the discharge~\cite{Pavlovic:1970}. In a high magnetic field (B $\gtrsim$1.5\,T), four pumping schemes can be used for the $2^3S_1$--$2^3P$ transitions (see Fig.~1 in Ref.~\cite{Nikiel:2007}), in this paper denoted as $f_2^\pm$ and $f_4^\pm$ where the subscript indicates the number of unresolved transition lines of the pumping scheme and + (-) represents the $\sigma^+$ right-handed ($\sigma^-$ left-handed) circular polarization of the 1083-nm pump light. For each pumping scheme, a separate pair of well resolved transition lines (probe doublet) of which the $2^3S_1$ sublevels are not addressed by the pumping lines can be used for optical polarimetry. In such a polarimetry approach, a probe laser is directed through the $^3$He with periodically sweeping wavelength over the probe doublet. The nuclear polarization of $^3$He $M$ is inferred by measuring the absorption signal amplitudes $a_1$ and $a_2$ ($a_1^0$ and $a_2^0$) for the probe doublet as the $^3$He is polarized (unpolarized, $M=0$), where
\begin{equation}    
\frac{a_2/a_1}{a_2^0/a_1^0}=\frac{1+M}{1-M},
\label{eq:polarimetry}
\end{equation}
the derivation of which can be found in Section 2 of ~\cite{Suchanek:2007}. Figure~\ref{fig:peaks} shows the measured absorption spectra for the $\sigma^+$ and $\sigma^-$ 1083-nm light at magnetic fields from 2 to 4\,T. The pump and probe peaks are subjected to Doppler broadening at room temperature and 1-torr pressure. Note that the degree of circular polarization of the 1083-nm light is not highly critical for the high-field MEOP as the $\sigma^+$ and $\sigma^-$ lines are well resolved due to the enhanced Zeeman splitting in high fields. 

\section{Experimental setup and method}
\label{sec:experiment}
\begin{figure}[htp]
\begin{center}
\includegraphics[width=1\columnwidth]{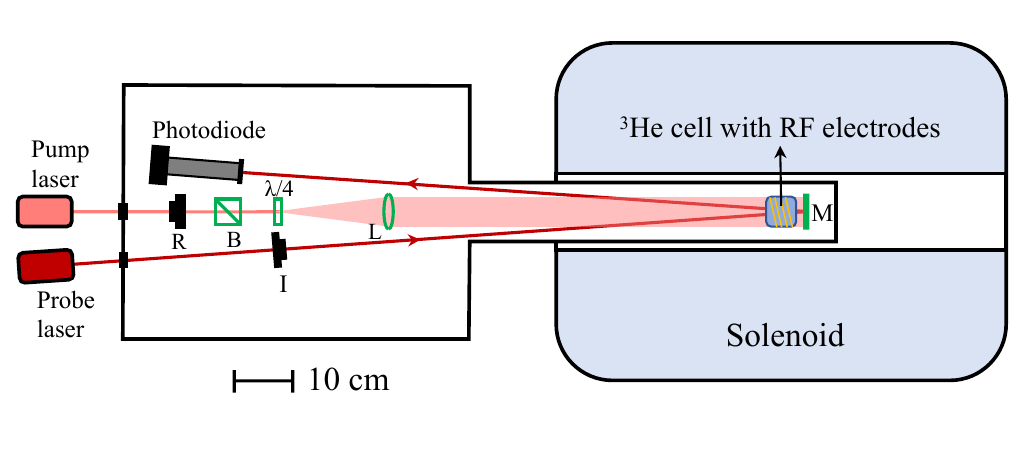}
\caption{Schematic of the experimental apparatus showing the layout of the optical pumping system and the probe polarimeter. Labeled optics components are fiber collimator on a rotatable mount (R), linearly polarizing beamsplitter cube (B), quarter-wave plate (λ/4), lens (L), fiber collimator coupled with an iris diaphragm (I) and mirror (M).
\label{fig:setup}}
\end{center}
\end{figure}

\begin{figure}[htp]
\begin{center}
\includegraphics[width=1\columnwidth]{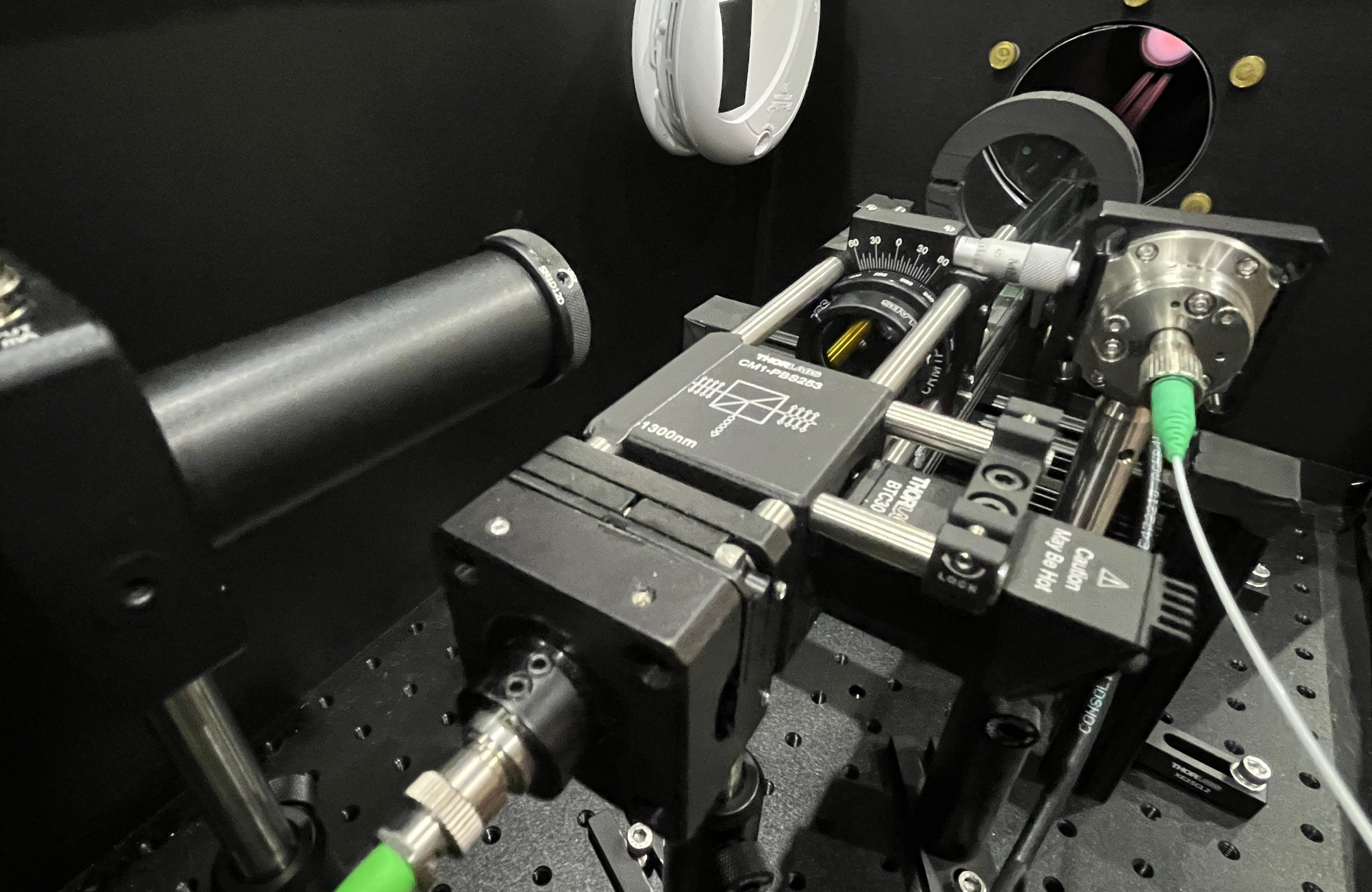}
\caption{Photograph of the optical bench inside the laser enclosure. The pump laser fiber is coupled to a collimator on a rotatable mount in the center to send the pumping light passing through a circular polarizer and a lens and incident on the $^3$He cell. The probe laser light is collimated on the right and passes through the cell before being reflected by a mirror and received by a photodiode on the left. A lens tube is coupled to the photodiode to suppress the background from laser reflection. The brightness of pink RF discharge plasma in the cell shown in the picture is much higher than in the actual optical pumping process.
\label{fig:optics}}
\end{center}
\end{figure}

The schematic layout of the experimental apparatus is shown in Fig.~\ref{fig:setup}. The $^3$He gas cell and all optical components are enclosed inside a laser-tight, rectangular enclosure (59\,cm in length, 43\,cm in width and 33\,cm in height) with a cylindrical extension (62\,cm in length and 10\,cm in diameter). Figure~\ref{fig:optics} shows a photograph of the optics bench within the laser enclosure box. The inner walls of the enclosure as well as all surfaces of the optical parts are darkened to minimize light reflection. The $^3$He glass cell is located near the end of the cylindrical volume, which is inserted into the warm bore of a superconducting magnet. The setup includes i) the optical pumping system consisting of the magnetic field, the $^3$He gas cell, the RF electrodes to produce discharge plasma in $^3$He gas and the pump laser and related optics, and ii) the optical polarimeter consisting of the probe laser, the mirror and the photodiode.

\subsection{Optical pumping system}
The superconducting magnet provides a homogeneous magnetic field up to 5\,T within the central area of its 76-cm-long and 13-cm-diameter warm bore. In this work, the magnet is operated at 2, 3 and 4\,T for the high-field MEOP tests of polarized $^3$He. Pure $^3$He gas of 1-torr pressure is sealed in a cylindrical borosilicate glass cell 5\,cm in length and 5\,cm in diameter. Electrical plasma discharge is induced by the electrodes spirally wound around the outer surface of the cell wall. An RF signal, typically between 10 and 50\,MHz, is generated by an SRS generator (SG382), amplified by an RF amplifier and tuned with a radio transformer before being sent to the electrodes. A Keopsys continuous-wave ytterbium-fiber laser system provides linearly polarized laser light with a tunable frequency range of about 100\,GHz, a nominal bandwidth of 2\,GHz and an output power range of 3--10\,W. The pumping light is delivered into the laser enclosure using a high-power optical fiber, which is coupled to a collimator on a rotatable mount inside the enclosure. The pumping light first passes through a linearly polarizing beamsplitter cube and a quarter-wave plate to ensure circular polarization, then guided by a lens to illuminate the full volume of the $^3$He cell.

\subsection{Optical polarimeter}
Taking advantage of the light absorption method introduced in~\cite{Suchanek:2007}, the optical polarimetry adopts the design as in~\cite{Maxwell:2018dyf} with slight modification. The probe laser light is provided by a Toptica laser system (DFB pro L-33508) with a tuning wavelength range of 1080.6--1084.2\,nm and an output power of 70\,mW. The probe light is guided by an optical fiber into the laser enclosure where it is split to be sent to a wavelength meter and the pumping cell. The cell fiber is coupled to a Thorlabs Fiberport collimator, after which the beam passes through an iris aperture to adjust the probe laser power delivered to the cell. 

\begin{figure}[htp]
\begin{center}
\includegraphics[width=1\columnwidth,center]{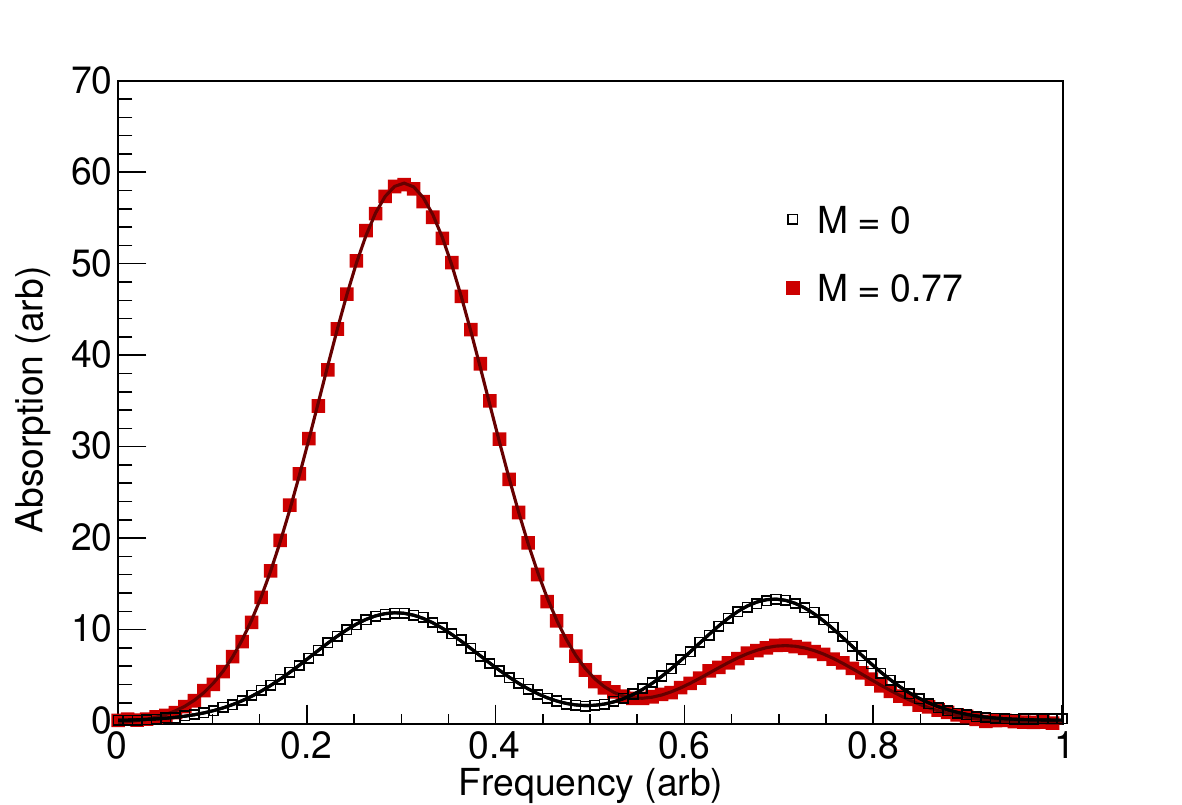}
\caption{Typical probe laser absorption signal measured with null (open squares) and 77\% (closed squares) nuclear polarizations using a 1\,Torr sealed cell at 2\,T. The solid curves are the Gaussian fits to the probe peaks.
\label{fig:polarimetry}}
\end{center}
\end{figure}

The frequency of the laser can be tuned by changing either the diode temperature or the operating current. The full range of probe laser frequency is explored by scanning the temperature to obtain the absorption spectrum for all pumping and probe peaks as shown in Fig.~\ref{fig:peaks}. The current is then swept over a smaller frequency range to map only the two absorption peaks of the probe doublet as shown in Fig.~\ref{fig:polarimetry}. The probe laser beam is incident on the cell at a small angle ($\sim$5\degree) with respect to the pumping light propagating direction, then reflected back through the cell by a broadband mirror (750--1100\,nm) downstream of the cell before finally reaching a photodiode (Thorlabs DET36A2) which is collimated with a lens tube to reduce the signal background from the reflected pumping light. 

To better isolate the probe signal received in the photodiode, the RF discharge is amplitude modulated from the SRS signal generator at 1\,kHz by 50\% modulation depth, which is taken as the reference for the lock-in amplifier (SRS SR860). The lock-in amplifier output is read to a computer using a Python program, where the measured spectrum for the probe doublet is fitted. The lineshape fit includes two side-by-side Gaussian peaks on top of a linear function to account for any background from the pump laser light as well as the linear shift in probe laser power caused by frequency sweeps. Figure~\ref{fig:polarimetry} shows examples of the measured probe doublet after the linear background subtraction for 0 and 77\% nuclear polarizations along with the Gaussian fitting curves. The absorption signal amplitudes $a_1$ and $a_2$ for the two probe peaks are extracted as the fitted amplitudes of the two Gaussian functions. The calibration measurement is taken at the beginning of each measurement cycle before the pump laser is turned on to obtain the absorption signal amplitudes $a_1^0$ and $a_2^0$ for null polarization. The nuclear polarization $M$ is determined from the measured value of $a_1$, $a_2$, $a_1^0$ and $a_2^0$ using Eq.~\ref{eq:polarimetry}.

\section{Results}
\label{sec:results}
\begin{figure}[!t]
\begin{center}
\includegraphics[width=1\columnwidth]{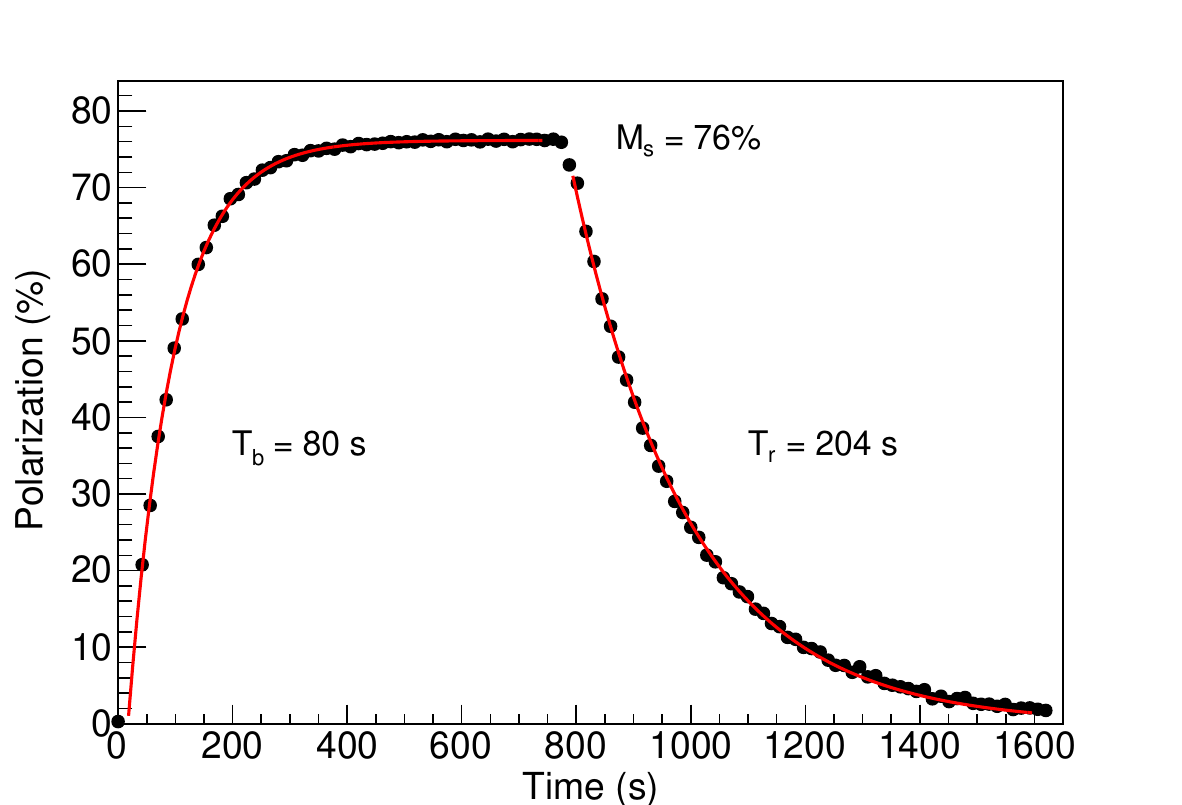}
\caption{Measurement cycle of optical pumping with $f_4^-$ scheme and discharge-on relaxation at 2\,T. The solid curves are the fits to the optical pumping and relaxation data. The build-up time $T_b$ and steady-state nuclear polarization $M_s$ are determined by fitting the optical pumping data from 0\,s to 770\,s with Eq.~\ref{eq:build-up}. The pump laser is off at 770\,s. The relaxation time $T_r$ is fitted as the exponential decay constant to the data beyond 770\,s.
\label{fig:cycle}}
\end{center}
\end{figure}

A typical optical pumping and relaxation cycle for the polarization measurement at 2\,T using the $f_4^-$ pumping scheme is shown in Fig.~\ref{fig:cycle}. 
The sweeps of the probe laser frequency for the probe doublet periodically and continuously proceed throughout the measurement cycle. 
Each polarization data point in Fig.~\ref{fig:cycle} is obtained from one full period of the sweep which typically takes about 14\,s. The pump laser is turned on at 0\,s and nuclear polarization of $^3$He starts to build up as an exponential function of time,  
\begin{equation}    
M(t) = M_s(1-e^{-\frac{t}{T_b}}),
\label{eq:build-up}
\end{equation}
where $M_s$ is the steady-state polarization and $T_b$ is the build-up time constant. Following the convention in~\cite{Gentile:1993}, the build-up rate, or effective pumping rate (pumping rate for short throughout this paper), $R$ is defined as
\begin{equation}    
R = \frac{NM_s}{T_b},
\label{eq:rate}
\end{equation}
where $N$ is the total number of atoms in the cell. Then the pump laser is turned off at 770 s to measure the relaxation process with the discharge on. The relaxation time $T_r$ is determined by fitting the relaxation data with an exponential decay function. 
$M_s$, $T_b$ and $T_r$ were measured with B-field magnitudes of 2, 3 and 4\,T to study the effects of the discharge intensity, pump laser power and different optical pumping transition schemes on high-field MEOP performance. The results will be presented and discussed in the following subsections.

\subsection{Discharge intensity}
The influence of the discharge condition on obtainable $^3$He nuclear polarization is two fold. On one hand, MEOP relies on the existence of the metastable-state $^3$He atoms which are produced by the RF discharge. The intensity of the discharge and its spatial distribution relative to the pump laser light directly determine the pumping rate.
On the other hand, discharge can lead to spin depolarization which is the major relaxation mechanism competing against the optical pumping process and hence affects the steady-state polarization of $^3$He nuclei. Generally, more intense discharge results in a stronger depolarization effect.
The distribution and intensity of the discharge are correlatively determined by the frequency and amplitude of the RF signal, the electrode configuration and the magnetic holding field.
The overall intensity of the discharge can be quantitatively controlled by varying the voltage amplitude of the RF signal and can be characterized by the relaxation time constant determined from discharge-on relaxation measurements. Longer relaxation time indicates weaker discharge within the cell. In this work, the discharge intensity was varied by fine tuning the output voltage of the RF generator and the transmitter. 

\begin{figure}[t]
\begin{center}
\includegraphics[width=1\columnwidth]{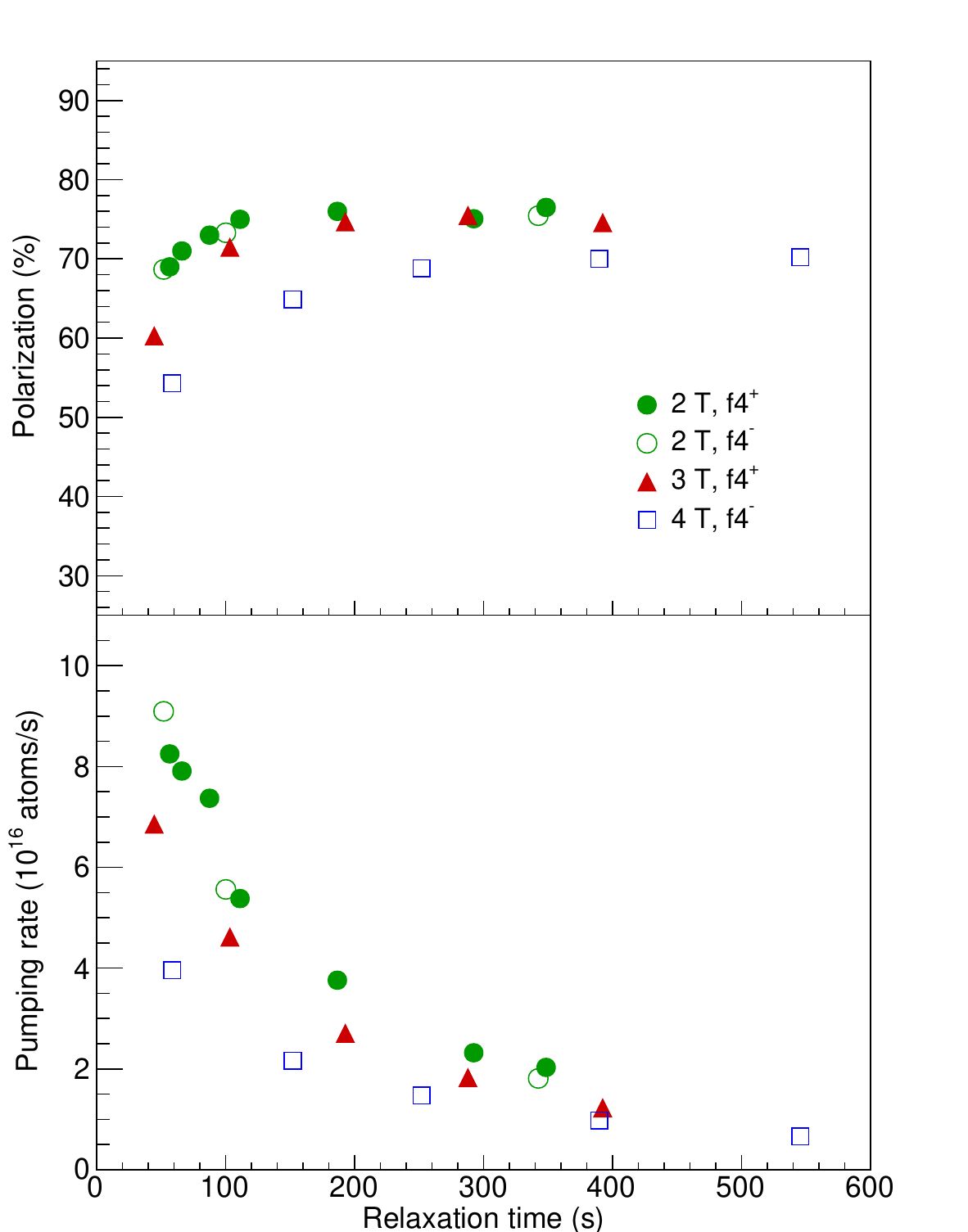}
\caption{Maximum steady-state nuclear polarization and pumping rate measured with various discharge intensity levels (characterized by the relaxation time) at 2\,T (circles), 3\,T (triangles) and 4\,T (squares) using the $f_4^\pm$ pumping schemes and a pump laser output power of 3\,W. The closed and open symbols are for the $\sigma^+$ and $\sigma^-$ schemes, respectively.
\label{fig:discharge}}
\end{center}
\end{figure}

Figure~\ref{fig:discharge} shows the steady-state nuclear polarization and the pumping rate measured at different discharge intensity levels, represented by the relaxation time, in magnetic fields of 2, 3 and 4\,T. The optical pumping was performed with the $f_4^\pm$ transition schemes and the output pump laser power was 3\,W. A saturation in the steady-state polarization is observed as the relaxation time prolongs. A trend of decreasing pumping rate and hence suppressed nuclear polarization with increasing magnetic field is obvious. The results are in reasonable agreement with those in~\cite{Maxwell:2018dyf}, which were obtained with the same $^3$He sealed cell 7 years prior. The discrepancy in saturation level could be rooted in the use of different superconducting magnets with different transverse B-field gradients and adoption of different electrode schemes which would in turn affect the discharge condition. To benchmark the depolarization effect resulting from the transverse field gradient, relaxation times with the discharge off were measured 2800--3800\,s.

\subsection{Pump laser power}
\begin{figure}[t]
\begin{center}
\includegraphics[width=1\columnwidth]{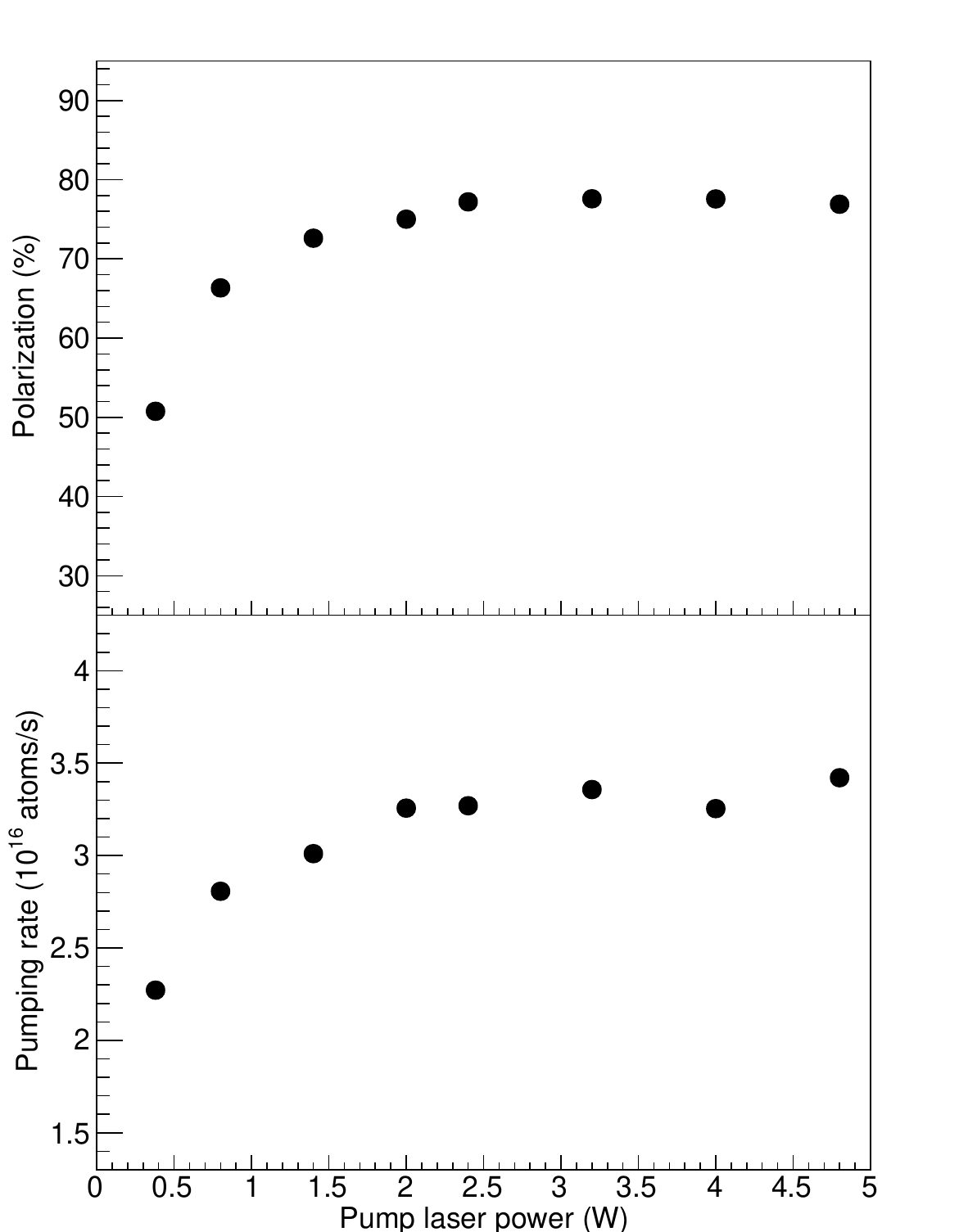}
\caption{Maximum steady-state nuclear polarization and pumping rate measured with various on-cell pump laser power at 2\,T using the $f_4^-$ pumping transition. The discharge-on relaxation time corresponding to each data point is above 200\,s.
\label{fig:power}}
\end{center}
\end{figure}

The influence of the pump laser power to the steady-state nuclear polarization and the pumping rate was evaluated. The setting range for the power output of the Keopsys laser is 3--10\,W. The laser light is attenuated by the linearly polarizing beamsplitter cube and the attenuation factor depends on the relative angle between the linear polarization plane of the output laser and that of the beamsplitter cube. By rotating the fiber mount around the propagation direction of the pump laser light, the on-cell laser power can be tuned between 0 and 3\,W. The actual power coming out of the quarter-wave plate was measured using a power meter as a function of the power supply setting and rotation angle of the fiber mount. Figure~\ref{fig:power} shows the dependence of the steady-state nuclear polarization and the pumping rate on laser power for the $f_4^-$ transition scheme at 2\,T. An on-cell power as low as about 2.5\,W is sufficient to reach the saturation of the attainable polarization.

\subsection{Pumping transition scheme}
\begin{figure}[]
\begin{center}
\includegraphics[width=1\columnwidth]{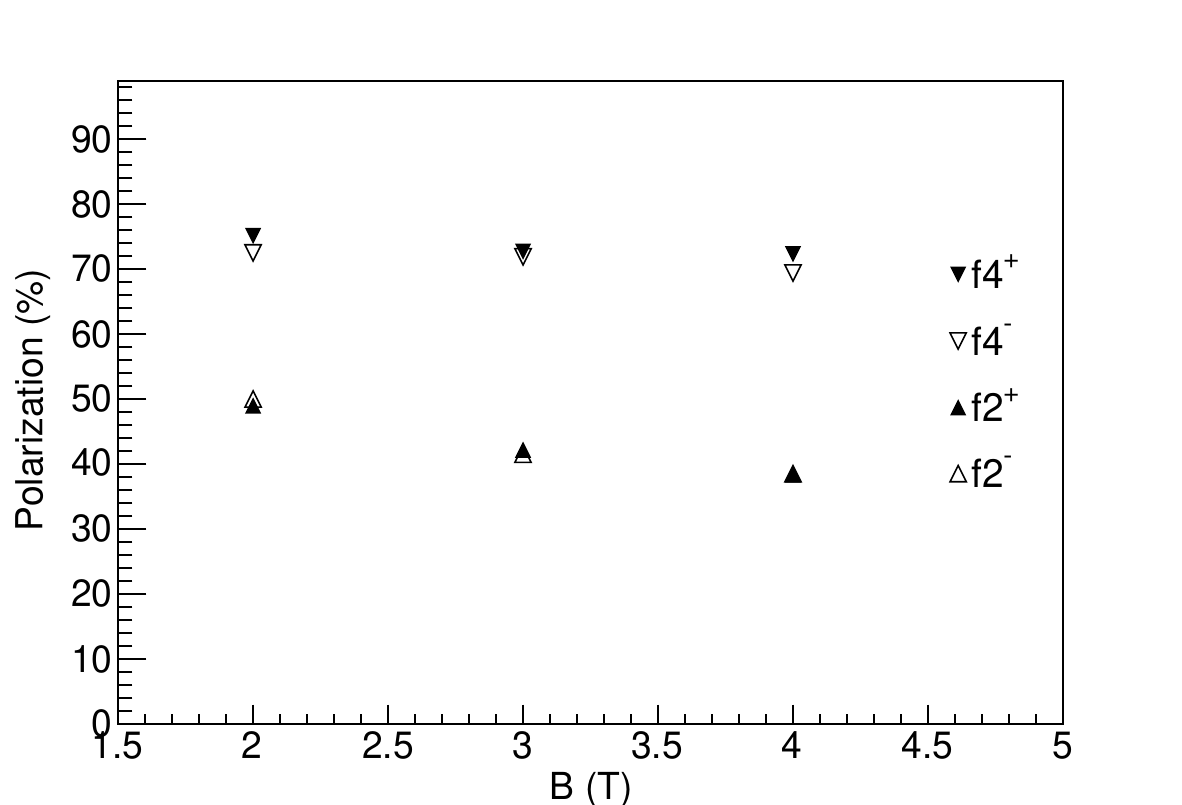}
\caption{Maximum steady-state nuclear polarization using four optical pumping schemes at 2 to 4\,T. Pump laser output power is 3\,W.
\label{fig:scheme}}
\end{center}
\end{figure}

Figure~\ref{fig:scheme} shows the maximum steady-state polarization achieved with the four optical pumping transition schemes described in Section~\ref{sec:meop}. The measurements were performed at 2, 3 and 4\,T with a pump laser power output of 3\,W. The results for all three magnetic fields consistently show that the $f_4^\pm$ schemes yield considerably higher nuclear polarization than the $f_2^\pm$ schemes. 
The $\sigma^+$ and $\sigma^-$ of the pumping light for either $f_2$ or $f_4$ schemes does not give apparent difference in the steady-state polarization taking into account the measurement uncertainties. The full results including the extracted pumping rate and relaxation time are tabulated in Table~\ref{tab:tab}.

\begin{table}[]
\begin{center}
\caption{Steady-state nuclear polarization ($M_s$), build-up time ($T_b$) and discharge-on relaxation time ($T_r$) measured at 2, 3 and 4\,T using the four optical pumping schemes and a pump laser output power of 3\,W. The pumping rate ($R$) is calculated using Eq.~\ref{eq:rate}.}
\begin{tabular*}{1\columnwidth}{@{\extracolsep{\fill}} c c  c c c c}
\noalign{\smallskip}
\hline\hline\noalign{\smallskip}
B & Transition  & $M_s$ & $T_b$ & $R$ & $T_r$\\
(T) & scheme  & (\%) & (s) & (atoms/s)& (s)\\
\noalign{\smallskip}\hline\noalign{\smallskip}
\multirow{4}{*}{2}& $f_2^+$ & 49  & 113 & 1.4$\times10^{16}$ & 278\\
                    & $f_4^+$ & 75  & 104 & 2.3$\times10^{16}$ & 293\\
                    & $f_2^-$ & 50  & 129 & 1.3$\times10^{16}$ & 297\\
                    & $f_4^-$ & 72  & 116 & 2.0$\times10^{16}$ & 286\\
\noalign{\smallskip}\hline\noalign{\smallskip}
\multirow{4}{*}{3}& $f_2^+$ & 42  & 193 & 7.0$\times10^{15}$ & 380\\
                    & $f_4^+$ & 70  & 194 & 1.2$\times10^{16}$ & 369\\
                    & $f_2^-$ & 41  & 196 & 6.7$\times10^{15}$ & 362\\
                    & $f_4^-$ & 70  & 193 & 1.2$\times10^{16}$ & 374\\
\noalign{\smallskip}\hline\noalign{\smallskip}
\multirow{4}{*}{4}& $f_2^+$ & 38  & 226 & 5.4$\times10^{15}$ & 388\\
                    & $f_4^+$ & 72  & 280 & 8.3$\times10^{15}$ & 379\\
                    & $f_2^-$ & 38  & 280 & 4.4$\times10^{15}$ & 375\\
                    & $f_4^-$ & 69  & 224 & 1.0$\times10^{16}$ & 404\\
\noalign{\smallskip}\hline\hline
\end{tabular*}
\label{tab:tab}
\end{center}
\end{table}

\subsection{Uncertainties}
The systematic uncertainties in the measurements for nuclear polarization mainly come from three aspects. 
i) The unsteadiness of the discharge light, which is dependent on the discharge level and the holding field, causes noise in the photodiode signal.
ii) The background light from the pump laser contributing to the photodiode signal which might not be completely addressed by the linear function in the fitting process.
Both i) and ii) lead to uncertainties in the extraction for the absorption signal amplitudes.
iii) Residual non-zero nuclear polarization might exist in the calibration runs for $a_1^0$ and $a_2^0$ which may result in a baseline offset in the measured nuclear polarization.
These three factors give total uncertainties of 2--4\% in the measured nuclear polarization $M$.
In addition, the selection of the data range for the exponential fitting introduces uncertainties in the extraction of $M_s$, $T_b$ and $T_r$, particularly prominent for $T_b$. The total uncertainties assigned for $M_s$, $T_b$ and $T_r$ are about 4\%, 5\% and 4\%, respectively. The resultant total uncertainty for $R$ is about 6\%, assuming that the uncertainty in the total number of atoms in the cell is negligible.

\section{Conclusion}
We present the first series of tests on MEOP at JLab for polarized $^3$He in high magnetic fields. The experiments have studied the dependence of discharge intensity and pump laser power for the attainable steady-state nuclear polarization and pumping rate, and indicated the optimal optical pumping schemes for the 1-torr gas. 
This work has reproduced and extended the earlier high-field MEOP results at BNL for the polarized $^3$He ion source for the EIC and serves as the baseline for the development of a novel polarized $^3$He gas target for CLAS12 at JLab.
The studies in the present work will be extended to 5\,T, the default B-field magnitude of the CLAS12 solenoid, using a new pump laser unit that provides a wider frequency range of 300\,GHz.
Future investigations, including polarization studies as a function of $^3$He gas pressure and magnetic field uniformity, and optimization of the discharge electrodes, will provide further input towards the design of a double-cell, cryogenic polarized $^3$He target for CLAS12.

\section*{Acknowledgments}
We thank the JLab Target Group for mechanical support. We are grateful for valuable discussions and support from Pierre-Jean Nacher and his colleagues at Laboratoire Kastler Brossel, Paris, France, and from Thomas Gentile at the National Institute of Standards and Technology, Gaithersburg, Maryland. We acknowledge the support of Nathan Isgur Fellowship. This research is supported by the U.S. Department of Energy Office of Nuclear Physics to the Massachusetts Institute of Technology under grant number DE-FG02-94ER40818 and to the Jefferson Lab under grant number DE-AC05-06OR23177.


\begin{thebibliography}{}
\bibitem{Colegrove:1960}
F.~D.~Colegrove, L.~D.~Schearer and G.~K.~Walters, 
Polarization of He$^3$ Gas by Optical Pumping, 
Phys. Rev. 132, 2561 (1963), 
\url{https://doi.org/10.1103/PhysRev.132.2561}.

\bibitem{Bouchiat:1960dsd}
M.~A.~Bouchiat, T.~R.~Carver and C.~M.~Varnum, 
Nuclear Polarization in He3 Gas Induced by Optical Pumping and Dipolar Exchange, 
Phys. Rev. Lett. 5, no.8, 373 (1960), 
\url{https://doi.org/10.1103/PhysRevLett.5.373}.

\bibitem{Jones:1993hg}
C.~E.~Jones, E.~J.~Beise, J.~E.~Belz, R.~W.~Carr, B.~W.~Filippone, W.~Lorenzon, R.~D.~McKeown, B.~A.~Mueller, T.~G.~O'Neill and G.~W.~Dodson, \textit{et al.} 
$^3\vv{\text{He}}\,(\vv{e}, e')$ quasielastic asymmetry, 
Phys. Rev. C 47, 110-130 (1993), 
\url{https://doi.org/10.1103/PhysRevC.47.110}.

\bibitem{Johnson:1994cq}
J.~R.~Johnson, A.~K.~Thompson, T.~E.~Chupp, T.~B.~Smith, G.~D.~Cates, B.~Driehuys, H.~Middleton, N.~R.~Newbury, E.~W.~Hughes and W.~Meyer, 
The SLAC high density gaseous polarized He-3 target, 
Nucl. Instrum. Meth. A 356, 148-152 (1995), 
\url{https://doi.org/10.1016/0168-9002(94)01465-5}.

\bibitem{DeSchepper:1998gc}
D.~DeSchepper, L.~H.~Kramer, S.~F.~Pate, K.~Ackerstaff, R.~W.~Carr, G.~R.~Court, A.~Dvoredsky, H.~Gao, A.~Golendoukhin and J.~O.~Hansen, \textit{et al.} 
The HERMES polarized He-3 internal gas target, 
Nucl. Instrum. Meth. A \textbf{419}, 16-44 (1998), 
\url{https://doi.org/10.1016/S0168-9002(98)00901-2}.

\bibitem{Krimmer:2009zz}
J.~Krimmer, M.~Distler, W.~Heil, S.~Karpuk, D.~Kiselev, Z.~Salhi and E.~W.~Otten, 
A highly polarized He-3 target for the electron beam at MAMI, 
Nucl. Instrum. Meth. A 611, 18-24 (2009), 
\url{https://doi.org/10.1016/j.nima.2009.09.064}.

\bibitem{Kramer:2007zzb}
K.~Kramer, X.~Zong, D.~Dutta, H.~Gao, X.~Qian, Q.~Ye, X.~Zhu, R.~Lu, T.~Averett and S.~Fuchs, 
A high-pressure polarized He-3 gas target for the High Intensity Gamma Source (HI$\gamma$S) facility at Duke Free Electron Laser Laboratory, 
Nucl. Instrum. Meth. A 582, 318-325 (2007), 
\url{https://doi.org/10.1016/j.nima.2007.08.243}.

\bibitem{Singh:2010}
J.~Singh, 
Alkali-Hybrid Spin-Exchange Optically-Pumped Polarized $^3$He Targets
Used For Studying Neutron Structure, 
Ph.D. thesis, University of Virginia (2010),
\url{http://galileo.phys.virginia.edu/research/groups/spinphysics/thesis/singh_thesis_2010.pdf}.

\bibitem{Milner:1989}
R.~G.~Milner, R.~D.~McKeown and C.~E.~Woodward, 
A polarized $^3$He target for nuclear physics, 
Nucl. Instrum. Meth. A 274, 56-63 (1989), 
\url{https://doi.org/10.1016/0168-9002(89)90365-3}.

\bibitem{Gao:1994ud}
H.~Gao, J.~Arrington, E.~J.~Beise, B.~Bray, R.~W.~Carr, B.~W.~Filippone, A.~Lung, R.~D.~McKeown, B.~Muller and M.~L.~Pitt, et al. 
``Measurement of the neutron magnetic form-factor from inclusive quasielastic scattering of polarized electrons from polarized He-3,'' 
Phys. Rev. C \textbf{50}, R546-R549 (1994), 
\url{https://doi.org/10.1103/PhysRevC.50.R546}.

\bibitem{Jones thesis}
C.~E.~Jones, Ph.D. thesis, California Institute of Technology, 1992.

\bibitem{Gao thesis}
H.~Gao, Ph.D. thesis, California Institute of Technology, 1994.

\bibitem{Courtade:2000}
E.~Courtade, F.~Marion, P.~Nacher, G.~Tastevin, T.~Dohnalik and K.~Kiersnowski, 
Spectroscopy of the helium 2 $^3$S–2 $^3$P transition above 0.01 tesla – application to optical pumping studies, 
Hyperfine Interact. 127 (1) 451–454 (2000), 
\url{https://doi.org/10.1023/A:1012673902661}.

\bibitem{Courtade:2002}
E.~Courtade, F.~Marion, P.-J.~Nacher, G.~Tastevin, K.~Kiersnowski and T.~Dohnalik, 
Magnetic field effects on the 1083 nm atomic line of helium, 
Eur. Phys. J. D 21, 25–55 (2002), 
\url{https://doi.org/10.1140/epjd/e2002-00176-1}.

\bibitem{Abboud:2004}
M.~Abboud, A.~Sinatra, X.~Maître, G.~Tastevin and P.-J.~Nacher, 
High nuclear polarization of $^3$He at low and high pressure by metastability exchange optical pumping at 1.5 tesla, 
Europhys. Lett. 68 (4) 480–486 (2004), 
\url{https://doi.org/10.1209/epl/i2004-10237-y}.

\bibitem{Abboud:2005}
M.~Abboud, A.~Sinatra, G.~Tastevin, P.-J.~Nacher and X.~Maître, 
Metastability Exchange Optical Pumping of Helium-3 at High Pressures and 1.5 T: Comparison of two Optical Pumping Transitions, 
\href{https://doi.org/10.48550/arXiv.physics/0506044}{arXiv:physics/0506044}.

\bibitem{Nikiel:2007}
A.~Nikiel, T.~Palasz, M.~Suchanek, M.~Abboud, A.~Sinatra, Z.~Olejniczak, T.~Dohnalik, G.~Tastevin and P.-J.~Nacher, 
Metastability exchange optical pumping of $^3$He at high pressure and high magnetic field for medical applications, 
Eur. Phys. J. Spec. Top. 144, 255–263 (2007), 
\url{https://doi.org/10.1140/epjst/e2007-00138-3}.

\bibitem{Suchanek:2007}
K.~Suchanek, M.~Suchanek, A.~Nikiel, T.~Pałasz, M.~Abboud, A.~Sinatra, P.-J.~Nacher, G.~Tastevin, Z.~Olejniczak and T.~Dohnalik, 
Optical measurement of $^3$He nuclear polarization for metastable exchange optical pumping studies at high magnetic field, 
Eur. Phys. J. Spec. Top. 144 (1) 67–74 (2007), 
\url{https://doi.org/10.1140/epjst/e2007-00109-8}.

\bibitem{Nikiel-Osuchowska:2013}
A.~Nikiel-Osuchowska, G.~Collier, B.~Głowacz, T.~Pałasz, Z.~Olejniczak, W.~P.~Wglarz, G.~Tastevin, P.-J.~Nacher and T.~Dohnalik, 
Metastability exchange optical pumping of $^3$He gas up to hundreds of millibars at 4.7 Tesla, 
Eur. Phys. J. D 67, 200 (2013), 
\url{https://doi.org/10.1140/epjd/e2013-40153-y}.

\bibitem{Zelenski:2023kof}
A.~Zelenski, G.~Atoian, E.~Beebe, S.~Ikeda, T.~Kanesue, S.~Kondrashev, J.~Maxwell, R.~Milner, M.~Musgrave and M.~Okamura, A.~A.~Poblaguev, D.~Raparia, J.~Ritter, A.~Sukhanov and S.~Trabocchi, 
Optically Pumped Polarized $^3$He$^{++}$ Ion Source Development for RHIC/EIC, 
Nucl. Instrum. Meth. A 1055, 168494 (2023),
\url{https://doi.org/10.1016/j.nima.2023.168494}.

\bibitem{Maxwell:2018dyf}
J.~D.~Maxwell, J.~Alessi, G.~Atoian, E.~Beebe, C.~S.~Epstein, R.~G.~Milner, M.~Musgrave, A.~Pikin, J.~Ritter and A.~Zelenski, 
Enhanced polarization of low pressure $^3$He through metastability exchange optical pumping at high field, 
Nucl. Instrum. Meth. A 959, 161892 (2020), 
\url{https://doi.org/10.1016/j.nima.2019.02.019}.

\bibitem{Maxwell:2021ytu}
J.~Maxwell and R.~Milner, 
A concept for polarized $^3$He targets for high luminosity scattering experiments in high magnetic field environments, 
Nucl. Instrum. Meth. A 1012, 165590 (2021), 
\url{https://doi.org/10.1016/j.nima.2021.165590}.

\bibitem{PAC:2020}
J.P.~Committee, \href{https://www.jlab.org/exp_prog/PACpage/PAC48/PAC48_PrelimReportPlus_FINAL.pdf}{48th program advisory committee report}, 2020.

\bibitem{Gentile:2016uud}
T.~R.~Gentile, P.~J.~Nacher, B.~Saam and T.~G.~Walker, 
Optically Polarized $^3$He, 
Rev. Mod. Phys. 89, no.4, 045004 (2017), 
\url{https://doi.org/10.1103/RevModPhys.89.045004}.

\bibitem{Pavlovic:1970}
M.~Pavlović and F.~Laloë, 
Study of a new method for orienting excited atomic levels by optical pumping. Application to the measurement of the hyperfine structure of 1D levels of $^3$He, 
J. Phys. France 31, 173-194 (1970), 
\url{http://dx.doi.org/10.1051/jphys:01970003102-3017300}.

\bibitem{Gentile:1993}
T.~R.~Gentile and R.~D.~McKeown, 
Spin-polarizing $^3$He nuclei with an arc-lamp-pumped neodymium-doped lanthanum magnesium hexaluminate laser, 
Phys. Rev. A 47 456–467 (1993), 
\url{http://dx.doi.org/10.1103/PhysRevA.47.456}.


\end{thebibliography}
\end{document}